\documentclass{ws-ijmpcs}

\begin{document}

\markboth{Shin'ichi Nojiri}
{Covariant gravity with Lagrange multiplier constraint}

%
\catchline{}{}{}{}{}
%

\title{COVARIANT GRAVITY@WITH LAGRANGE MULTIPLIER CONSTRAINT
}

\author{SHIN'ICHI NOJIRI}

\address{Department of Physics, Nagoya University, Nagoya
464-8602, Japan \\
and Kobayashi-Maskawa Institute for the Origin of Particles and
the Universe, Nagoya University, Nagoya 464-8602, Japan 
}



\maketitle

\begin{history}
\received{Day Month Year}
\revised{Day Month Year}
\end{history}

\begin{abstract}
We review on the models of gravity with a constraint by the 
Lagrange multiplier field. The constraint breaks general covariance or 
Lorentz symmetry in the ultraviolet region. 
We report on the $F(R)$ gravity model with the 
constraint and the proposal of the covariant (power-counting) 
renormalized gravity model by using the constraint and scalar projectors.
We will show that the model admits flat space solution, 
its gauge-fixing formulation is fully developed, and the only propagating 
mode is (higher derivative) graviton, while scalar and vector modes 
do not propagate.
The preliminary study of FRW cosmology indicates to
the possibility of inflationary universe solution is also given.

\end{abstract}

\ccode{PACS numbers: 11.25.Hf, 123.1K}

\section{Introduction}	

Recently the models of gravity with a constraint by the 
Lagrange multiplier field have been 
proposed\cite{vikman,Capozziello:2010uv,Nojiri:2009th,Kluson:2011rs}. 
The constraint breaks general covariance or Lorentz symmetry and given by 
the following action: 
\begin{align}
\label{Q1}
S = - \int d^4 x \sqrt{-g} \lambda \left( \frac{1}{2} \partial_\mu \phi \partial^\mu \phi
+ U_0 \right) \, .
\end{align}
Here $\lambda$ is the Lagrange multiplier field and $U_0$ is a positive constant. 
Then the variation of $\lambda$, we obtain the following equation: 
\begin{align}
\label{Q2}
\frac{1}{2} \partial_\mu \phi \partial^\mu \phi
+ U_0 = 0\, ,
\end{align}
which tells $(\partial_\mu \phi)$ is a non-vanishing time-like vector and 
therefore the general covariance or the Lorentz symmetry is breakdown 
spontaneously. 
Locally, one can choose the direction of time to be parallel to $(\partial_\mu \phi)$. 
Then we find 
\begin{align}
\label{Q37}
\phi = \sqrt{2 U_0} t\, ,
\end{align}

In this report, especially we review on the $F(R)$ gravity model with the 
constraint\cite{Capozziello:2010uv} and the proposal of the covariant (power-counting) 
renormalized gravity model\cite{Nojiri:2009th} 
by using the constraint and scalar projectors\cite{Kluson:2011rs}.

In 2009, Ho\v{r}ava proposed a  candidate of quantum field theory of gravity 
which is power-counting renormalizable. 
The model has anisotropy between space and time by the explicit breaking of covariance 
but it was expected that at long distances, the Lorentz symmetry could be recovered. 
It was clarified, however, the existence of extra scalar mode violating the Newton 
law\cite{Charmousis:2009tc,Li:2009bg,Blas:2009qj}. 
After that there were proposals of covariant and power-counting renormalizable model 
of gravity\cite{Nojiri:2009th,Kluson:2011rs}. It has been shown show that 
only massless graviton propagates in the model of Ref.~\refcite{Kluson:2011rs}. 

\section{Application to $F(R)$ gravity}

As an application of the Lagrange multiplier field to the $F(R)$ gravity models, 
we consider the following action:
\begin{align}
\label{Q4} 
S = \int d^4 x \sqrt{-g} \left\{ F_1(R) - \lambda
\left( \frac{1}{2}
\partial_\mu R \partial^\mu R
+ F_2 (R) \right) \right\}\, .
\end{align}
By the variation of $\lambda$, we obtain the following constraint
\begin{align}
\label{Q5}
\frac{1}{2} \partial_\mu R \partial^\mu R + F_2(R) = 0\, .
\end{align}
On the other hand, by the variation of the metric $g_{\mu\nu}$, we obtain
\begin{align}
\label{Q6}
0 =& \frac{1}{2} g_{\mu\nu} F_1(R) + \frac{\lambda}{2} \partial_\mu R \partial_\nu R
+ \left( - R_{\mu\nu} + \nabla_\mu \nabla_\nu - g_{\mu\nu} \nabla^2 \right) \nonumber \\
& \times \left( F_1'(R) - \lambda F_2' (R) - \nabla^\mu \left(\lambda
\nabla_\mu R \right) \right)\, .
\end{align}
If the Ricci curvature is covariantly constant: $R_{\mu\nu} = \frac{R_0}{4}g_{\mu\nu}$ 
and therefore $R=R_0$, 
Eqs.~(\ref{Q5}) and (\ref{Q6}) give $0 = F_2 (R_0)$ and 
$0 = F_1(R_0) - \frac{1}{2} R_0 \left( F_1' (R_0) - \lambda F_2' (R_0) \right)$ 
and therefore $\lambda = \frac{ - 2 F_1 (R_0) + R_0 F_1'(R_0)}{R_0 F_2'(R_0)}$.
Then if $R_0>0$, we obtain a solution describing de Sitter space-time, which may be 
regarded with the inflation. 

For spatially-flat FRW metric
\begin{align}
\label{Q10}
ds^2 = - dt^2 + a(t)^2 \sum_{i=1,2,3} \left(dx^i\right)^2\, .
\end{align}
Eqs.~(\ref{Q5}) and (\ref{Q6}) have the following form:
\begin{align}
\label{Q11}
0 =& - \frac{1}{2}{\dot R}^2 + F_2(R) \, ,\nonumber \\
0 =& - \frac{1}{2} F_1(R) + 18 \lambda \left(\ddot H + 4 H \dot H \right)^2
+ \left\{3\left( \dot H + H^2 \right) - 3H\frac{d}{dt} \right\} \nonumber \\
& \times \left\{ F_1'(R) - \lambda F_2' (R) + \left(\frac{d}{dt} + 3H \right) \left( \lambda
\frac{dR}{dt} \right) \right\} \, .
\end{align}
By integrating the first equation in (\ref{Q11}) when $F_2(R)>0$, we obtain $t = \int^R \frac{dR}{\sqrt{2F_2(R)}}$, 
which can be solved with respect to $R$ as $R=F_R(t)$. Then since
$6 \frac{dH}{dt} + 12 H^2 = R  =F_R(t)$, 
we obtain the $t$ dependence of $H$ and $\lambda$ as $H=H(t)$ and $\lambda=\lambda(t)$. 

Conversely if we know $H(t)$, we can construct $F_2(R)$ to produce the $t$ dependence of $H$. 
As we know $H(t)$, we can find the $t$ dependence of the scalar curvature $R=R(t)$, which 
could be solved with respect to $t$ as $t=t(R)$. 
Then the first equation in (\ref{Q11}) gives the explicit form of $F_2(R)$ as
$F_2(R) = \frac{1}{2}\left. \left(\frac{d R}{dt}\right)^2 \right|_{t=t(R)}$. 
We should note $F_1(R)$ can be an arbitrary function.

As an example, we may consider $H(t) = \frac{h_0}{t}$. 
Here $h_0$ is a positive constant. Then we find 
$F_2(R) = \frac{R^3}{12 \left( - h_0 + 2 h_0^2 \right) }$. 
As a second example, we may consider
\begin{align}
\label{Q16}
R = \frac{R_-}{2} \left( 1 - \tanh \omega t \right)
+ \frac{R_+}{2} \left( 1 + \tanh \omega t \right)\, .
\end{align}
Here $R_\pm$ and $\omega$ are positive constants. For the curvature $R$ in 
(\ref{Q16}), we find $R\to R_\pm$ when $t\to \pm \infty$, that is, 
asymptotically de Sitter space-time. Then we may regard $t\to -\infty$ 
could correspond to inflation, and $t\to +\infty$ to late acceleration.
We now find
\begin{align}
\label{Q17}
F_2(R) = \frac{\left(R_- - R_+\right)^2 \omega^2}{8} \left( 1
   - \frac{\left( R_- + R_+ - 2R \right)^2}{\left( R_- - R_+ \right)^2}
\right)^2\, .
\end{align}

\section{Brief review on Ho\v{r}ava gravity}

Before we apply the constraint by the Lagrange multiplier field 
to the covariant power-counting renormalizable gravity, 
we will briefly review the model in Ref.~\refcite{Horava:2009uw}. 
We know that Einstein's general relativity is 
non-renormalizable as a quantum field theory. If we consider the 
perturbation from the flat background: $g_{\mu\nu} = \eta_{\mu\nu} + \kappa h_{\mu\nu}$ 
(Here $\kappa$ is a gravitational coupling), the Einstein-Hilbert action has the 
following form: 
\begin{align}
\label{Q18}
S =& \frac{1}{2\kappa^2} \int d^4 x \sqrt{-g} R \nonumber \\
=& \int d^4 x \left[ - \frac{1}{2} \partial h \partial h + \kappa h \partial h \partial h 
+ \kappa^2 h^2 \partial h \partial h + \cdots \quad + \kappa^n h^n \partial h \partial h + \cdots 
\right]\, .
\end{align}
Since the dimension of $\kappa$ is that of length, Einstein's general relativity 
is non-renormalizable. 
However, if the propagator behaves as $1/p^4$, $1/p^6$, $\cdots$ instead of $1/p^2$ 
($p_\mu$: four momentum) in the ultraviolet region, the ultraviolet behavior 
of the quantum correction could be improved. Usually such a propagator is given by 
the higher derivative theory, where the unitarity could be broken in general, 
due the higher derivative with respect to $t$

Then Ho\v{r}ava's idea is to introduce anisotropic treatment between space and time
and consider the higher derivative theory with respect to only spacial coordinates. 
Then in the ultraviolet region, the propagator behaves as $1/\bm{p}^4$, $1/\bm{p}^6$, $\cdots$. 
The anisotropy an be expressed by a parameter $z$ which is given by the scale 
transformation with a constant $b$: $\bm{x}\to b\bm{x}$, $t\to b^z t$, ($z=2,3,\cdots$). 

In order to express the action of the model in Ref.~\refcite{Horava:2009uw}, 
we use the well-known ADM decomposition, where the metric is expressed as 
\begin{align}
\label{Q19}
ds^2 = - N^2 dt^2 + \sum_{i,j=1,2,3}g^{(3)}_{ij}\left(dx^i + N^i dt \right)\left(dx^j + N^j dt \right) \, .
\end{align}
Here $N$ and $N^i$ are called the lapse variable and the shift variable, respectively. 
Then we find that the Einstein-Hilbert action has the form of 
$\int d^3 x dt N \sqrt{g^{(3)}} \left( K^{ij} K_{ij} - K^2 + R^{(3)}\right)$. 
Here $K_{ij}$ is the extrinsic curvature defined by
$K_{ij}=\frac{1}{2N}\left(\dot g_{ij}-\nabla_iN_j-\nabla_jN_i\right)$ 
and $K =K^i_{\ i}$. 
The action of Ho\v{r}ava gravity is given by a sum of ``kinetic term'' and ``potential'' 
The kinetic term is the term including the derivatives with respect to time and given by
\begin{align}
\label{Q22}
S_K=\frac{2}{\kappa^2}\int dt\,d^3 \bm{x}\,\sqrt{g}N\left(K_{ij}K^{ij}
-\lambda K^2\right)\, ,
\end{align}
Here $\lambda$ is a parameter.\footnote{
Here we followed the notation in Ref.~\refcite{Horava:2009uw}. Except in this section, 
$\lambda$ expresses the Lagrange multiplier field.
}
The action is invariant under the spacial diffeomorphism and the temporal diffeomorphism, 
which are given by
\begin{align}
\label{Q23}
\delta x^i=\zeta^i(t,\bm{x})\, ,\qquad \delta t=f(t)\, .
\end{align}
By using $z$, the dimension of time $t$ is expressed as $[L^z]$. 
Here $L$ is the length. Since $ds^2 = - N^2 dt^2 + \cdots$, 
we find the dimension of $N$ as $[N] = [L^{1-z}]$. 
In perturbation theory, when we fix the gauge of diffeomorphism, we choose $N=N_0$
with a constant $N_0$. 
Then as clear from (\ref{Q22}), the effective coupling constant is given by 
$1/\kappa_\mathrm{eff}^2 = N_0/\kappa^2 $, whose dimension is  
$[ \kappa_\mathrm{eff}^2 ] = [ L^{3 - z} ]$. 
Then when $z=3$, $\kappa_\mathrm{eff}$ becomes dimensionless and therefore 
the model becomes power counting renormalizable. 

\newcommand{\lambdaw}{\Lambda_W^{}}
\newcommand{\bx}{\bm{x}}
\newcommand{\CG}{\mathcal{G}}
\newcommand{\trace}{\mathrm{Tr}}
\newcommand{\p}{\partial}

In order to include the ``potential'', that is, the terms not including 
the derivatives with respect to time, the generalized De~Witt ``metric 
on the space of metrics'', which is given by 
$\CG^{ijk\ell}=\frac{1}{2}\left(g^{ik}g^{j\ell}+g^{i\ell}g^{jk}\right)-\lambda 
g^{ij}g^{k\ell}$. 
Then Ho\v{r}ava has proposed the potential with ``detailed balance'', 
which is given by
\begin{align}
\label{Q25}
S_V=\frac{\kappa^2}{8}\int dt\,d^3 \bx\,\sqrt{g}N\,E^{ij}\CG_{ijk\ell} 
E^{k\ell}\ , \quad 
\sqrt{g}E^{ij}=\frac{\delta W[g_{k\ell}]}{\delta g_{ij}}
\end{align}
Then for the $z=2$ model, Ho\v{r}ava has chosen 
$W=\frac{1}{\kappa_W^2}\int d^3 \bx\,\sqrt{g}(R-2\lambdaw)$, 
which gives
\begin{align}
\label{Q27}
S_V=\frac{\kappa^2}{8\kappa_W^4}\int dt\,d^3 \bx\,\sqrt{g}N
\left(R^{ij}-\frac{1}{2}Rg^{ij}+\lambdaw g^{ij}\right) 
\CG_{ijk\ell} 
\left(R^{k\ell}-\frac{1}{2}Rg^{k\ell}+\lambdaw g^{k\ell}\right)\, .
\end{align}
For $z=3$ model, it has proposed as 
$W=\frac{1}{w^2}\int_\Sigma\omega_3(\Gamma)$. 
Here $w^2$ is a dimensionless coupling and 
$\omega_3(\Gamma)$ is gravitational Chern-Simons term given by
\begin{align}
\label{Q29}
\omega_3(\Gamma) = \trace\left(\Gamma\wedge d\Gamma+\frac{2}{3}\Gamma\wedge\Gamma
\wedge\Gamma\right) 
\equiv 
\varepsilon^{ijk}\left(\Gamma^{m}_{i\ell}\p_j
\Gamma^{\ell}_{km}+\frac{2}{3}\Gamma^{n}_{i\ell}\Gamma^{\ell}_{jm}
\Gamma^{m}_{kn}\right)d^3\bx \, .
\end{align}
Then we find
\begin{align}
\label{Q30}
S =& \int dt\,d^3\bx\,\sqrt{g}\,N\left\{\frac{2}{\kappa^2}\left(
K_{ij}K^{ij}-\lambda K^2\right)-\frac{\kappa^2}{2w^4}C_{ij}C^{ij}
\right\} \nonumber \\ 
=& \int dt\,d^3\bx\,\sqrt{g}\,N\left\{\frac{2}{\kappa^2}\left(
K_{ij}K^{ij}-\lambda K^2\right)\right. \nonumber \\ 
& \left. -\frac{\kappa^2}{2w^4}\left(\nabla_iR_{jk}
\nabla^iR^{jk}-\nabla_iR_{jk}\nabla^jR^{ik}-\frac{1}{8}\nabla_i R\nabla^i R
\right)\right\} \, .
\end{align}
Here $C^{ij}$ is called the Cotton tensor defined by
$C^{ij}=\varepsilon^{ik\ell}\nabla_k\left(R^j_\ell-\frac{1}{4}R\delta^j_\ell
\right)$. 

After the proposal, it has been clarified that there are several 
problems\cite{Charmousis:2009tc,Li:2009bg,Blas:2009qj} in the Ho\v{r}ava gravity. 
The model does not have full diffeomorphism symmetry but the direct product of 
the spacial diffeomorphism and temporal diffeomorphism (\ref{Q23}). 
Then in order to impose a gauge condition $N=$constant, we may assume the 
projectability condition, that is, $N$ should depend on only time coordinate $N=N(t)$. 
Although there have been also proposed models which do not satisfy the projectability condition, 
the degrees of freedom of the Ho\v{r}ava gravity do not coincide with those of 
the Einstein gravity and therefore the Ho\v{r}ava gravity does not reproduce general relativity 
even in the low energy region. 

\section{Proposal of covariant and power-counting renormalizable models of gravity}

In order to construct models with correct degrees of freedom, 
we have proposed a model with the covariance (full diffeomorphism invariance), 
that is, the covariant and power-counting renormalizable 
models\cite{Nojiri:2009th,Kluson:2011rs} by using the spontaneous breakdown 
of Lorentz symmetry. 
 
The action of the proposed model is given by
\begin{align}
\label{Q33}
& S = \int d^4 x \sqrt{-g} \left[ \frac{R}{2\kappa^2} - \alpha
\Bigl\{
\left(\partial^\mu \phi \partial^\nu \phi \nabla_\mu
\nabla_\nu - \partial_\mu \phi \partial^\mu \phi
\nabla^\rho \nabla_\rho \right)^n \right. P_\alpha^{\ \mu} P_\beta^{\ \nu} \nonumber \\ 
& \left. \times \left( R_{\mu\nu} - \frac{1}{2 U_0 }
\partial_\rho \phi \nabla^\rho \nabla_\mu \nabla_\nu \phi
\right) \right\} \Bigl\{
\left(\partial^\mu \phi \partial^\nu \phi \nabla_\mu
\nabla_\nu - \partial_\mu \phi \partial^\mu \phi
\nabla^\rho \nabla_\rho \right)^{n+\Delta}
P^{\alpha\mu} P^{\beta\nu} \nonumber \\ 
& \left. \left. \times \left( R_{\mu\nu} - \frac{1}{2 U_0 }
\partial_\rho \phi \nabla^\rho \nabla_\mu \nabla_\nu \phi
\right) \right\} - \lambda \left( \frac{1}{2} \partial_\mu \phi \partial^\mu \phi
+ U_0 \right) \right]\, ,
\end{align}
Here $\lambda$ is the Lagrange multiplier field, $U_0$ is a constant, and 
$P_\mu^{\ \nu}$ is a projection operator defined by 
$P_\mu^{\ \nu} \equiv \delta_\mu^{\ \nu} + \frac{\partial_\mu \phi
\partial^\nu \phi}{2U_0}$. For $z=2n + 2$ model $\left( n=0,1,2,\cdots\right)$, 
$\Delta=0$ and for $z=2n + 3$ model, $\Delta=1$. 
First we should note that the actions admit a flat space vacuum solution. 
The field equations have the following form:
$0 = \frac{1}{2\kappa^2} \left( R_{\mu\nu} - \frac{1}{2} g_{\mu\nu} R \right)
+ G^\mathrm{higher}_{\mu\nu} - \frac{\lambda}{2}
\partial_\mu \phi \partial_\nu \phi + \frac{\lambda}{2}
g_{\mu\nu}
\left( \frac{1}{2} \partial_\rho \phi \partial^\rho \phi + U_0 \right)$. 
Then by assuming the flat vacuum solution, we find 
$0 = \lambda \partial_\mu \phi \partial_\nu \phi$, 
which gives $\lambda=0$. Then we obtain the flat space
vacuum solution with $\lambda=0$.

\section{Perturbation from the flat background}

In this section, we consider the perturbation from the flat background and 
we show that the only propagating mode is
higher derivative graviton while scalar and vector modes do not
propagate. 

We now fix the diffeomorphism invariance with respect to time 
coordinate by choosing the condition by (\ref{Q37}), 
which is a kind of the unitary gauge condition. 
Then by the perturbation from flat background: 
$g_{\mu\nu} = \eta_{\mu\nu} + h_{\mu\nu}$, we find 
\begin{align}
\label{Q38}
& S \to \int d^4 x \left[ - \frac{1}{8\kappa^2} \left\{ - 2 h_{tt}
\left( \delta^{ij} \partial_k \partial^k - \partial^i \partial^j
\right) h_{ij} + 2 h_{ti} \left( \delta^{ij} \partial_k
\partial^k - \partial^i \partial^j \right) h_{tj}
\right. \right. \nonumber \\ 
& + h_{ti} \left( 2 \delta^{jk} \partial^i - \delta^{ik}
\partial^j - \delta^{ij} \partial^k \right) \partial_t h_{jk} 
+ h_{ij} \left( \left( \delta^{ij}
\delta^{kl} - \frac{1}{2} \delta^{ik} \delta^{jl} - \frac{1}{2}
\delta^{il} \delta^{jk} \right)
\left( - \partial_t^2 + \partial_k \partial^k\right) \right. \nonumber \\ 
& \left.\left. - \delta^{ij} \partial^k \partial^l - \delta^{kl}
\partial^i \partial^j
+ \frac{1}{2} \left( \delta^{ik} \partial^j \partial^l + \delta^{il}
\partial^j \partial^k
+ \delta^{jk} \partial^i \partial^l + \delta^{jl} \partial^i \partial^k
\right) \right) h_{kl} \right\} \nonumber \\ 
& - 2^{2n-2+\Delta} \alpha U_0^{2n+\Delta} \left\{ \left( \partial_k \partial^k \right)^n
\left(h_{ki,j}^{\ \ \ \ k}
+ h_{kj,i}^{\ \ \ \ k} - h_{ij,k}^{\ \ \ \ k} - \partial_i \partial_j
\left( h_\mu^{\ \mu} \right) \right) \right\} \nonumber \\ 
& \left. \times \left\{ \left( \partial_k \partial^k \right)^{n+\Delta}
\left(h_{k\ ,}^{\ i\ jk}
+ h_{k\ ,}^{\ j\ ik} - h_{\ \ ,k}^{ij\ \ k} - \partial^i \partial^j
\left( h_\mu^{\ \mu} \right) \right) \right\}
+ U_0 \lambda h_{tt} \right] \, .
\end{align}
By the variation of $\lambda$, we obtain $h_{tt}=0$. 
On the other hand by the variation of $h_{tt}$, we obtain
\begin{align}
\label{Q40}
\lambda =& - \frac{1}{4\kappa^2 U_0 } \left( \delta^{ij} \partial_k
\partial^k - \partial^i \partial^j \right) h_{ij} \nonumber \\ 
& + 2^{2n-1+\Delta} \alpha U_0^{2n-1+\Delta} \left( \partial_k \partial^k \right)^{2n+\Delta}
\partial^i \partial^j
\left(h_{ki,j}^{\ \ \ \ k}
+ h_{kj,i}^{\ \ \ \ k} - h_{ij,k}^{\ \ \ \ k} - \partial_i \partial_j
\left( h_\mu^{\ \mu} \right) \right) \, . 
\end{align}
The variation of $\phi$ gives
\begin{align}
\label{Q42}
0 =& \partial_t \left\{\lambda + 2^{2n - 1 +\Delta } \alpha U_0^{2n - 1 + \Delta}
\left( \partial_k \partial^k \right)^{2n+\Delta} \partial^i \partial^j
\left(h_{ki,j}^{\ \ \ \ k}
+ h_{kj,i}^{\ \ \ \ k} - h_{ij,k}^{\ \ \ \ k} - \partial_i \partial_j
\left( h_\mu^{\ \mu} \right) \right) \right\} \, .
\end{align}
We now decompose $h_{ti}$, which corresponds to the shift function $N_i$, 
as $h_{ti}= \partial_i s + v_i$. Here $v_i$ satisfies the equation 
$\partial^i v_i = 0$ and $s$ is a spatial scalar. 
We also linearize the diffeomorphism invariance transformations
with respect to the spatial coordinates as 
$\delta x^i = \partial^i u + w^i$, where $w_i$ satisfies the equation 
$\partial_i w^i = 0$. Then we find $\delta s = \partial_t u$ and 
$\delta v_i = \partial_t w_i$. 
We now choose the gauge fixing condition as $s=v^i=0$, which gives $h_{ti}=0$. 
The variation of $h_{ti}$ gives
\begin{align}
\label{Q46}
\partial_t \left( -2 \delta^{jk} \partial^i + \delta^{ik} \partial^j
+ \delta^{ij} \partial^k \right) h_{jk} = 0\, ,
\end{align}
which is identical with that in the Einstein gravity and does not include 
higher derivative terms. 

We now also decompose $h_{ij}$ as 
$h_{ij} = \delta_{ij} A + \partial_j B_i + \partial_i B_j + C_{ij} + \left(
\partial_i \partial_j - \frac{1}{3}\delta_{ij}
\partial_k \partial^k \right) E$, where $B_i$ and $C_{ij}$ satisfy equations 
$\partial^i B_i = 0$, $\partial^i C_{ij} = \partial^j C_{ij} = 0$, and 
$C_i^{\ i}=0$.
Then by the variation of $\delta h_{ti}$, we obtain
\begin{align}
\label{Q48}
0 = \partial_t \left( -4 \partial_i A + 2 \partial_k \partial^k B_i
+ \frac{4}{3}\partial_i \partial_k \partial^k E \right) \, .
\end{align}
By multiplying (\ref{Q48}) with $\partial^i$, we find
$\partial_t \partial_i \partial^i \left( - 4 A
+ \frac{4}{3} \partial_k \partial^k E \right) =0$, which gives 
$A = \frac{1}{3} \partial_k \partial^k E$. 
Here we have assumed $A$ and $E$ vanish at spatial infinity.
By using (\ref{Q48}), we find
$\partial_t \partial_j \partial^j B_i = 0$, which gives $B_i=0$ 
by assuming $B_i\to 0$ at spatial infinity.
Then Eqs.~(\ref{Q40}) and (\ref{Q42}) can be rewritten as 
\begin{align}
\label{Q50}
\lambda =& \frac{1}{2\kappa^2 U_0} \partial_k \partial^k \left( - A
+ \frac{1}{3} \partial_j \partial^j E \right) - 2^{2n+\Delta} \alpha U_0^{2n-1+\Delta}
\left( \partial_k \partial^k \right)^{2n + 2+\Delta}
\left( - A + \frac{1}{3} \partial_j \partial^j E \right)\, , \\
\label{Q51}
0 =& \partial_t \left\{ \lambda + 2^{2n+\Delta} \alpha U_0^{2n - 1 +\Delta}
\left( \partial_k \partial^k \right)^{2n+2+\Delta} \left( - A + \frac{1}{3}
\partial_j \partial^j E \right) \right\}\, .
\end{align}
Then we find $\lambda = 0$ since $A = \frac{1}{3} \partial_k \partial^k E$. 
Therefore we find the scalar modes $\lambda$ and the vector mode $B_i$ 
do not propagate.

By the variation of $A$ gives 
\begin{align}
\label{Q54}
0 =& \frac{1}{8\kappa^2} \left\{ - 12 \left( - \partial_t^2 + \partial_k
\partial^k\right) A
+ 8 \partial_k \partial^k A + \frac{4}{3} \left( \partial_k \partial^k
\right)^2 E \right\} \nonumber \\ 
& - 2^{2n-1+\Delta} \alpha U_0^{2n+\Delta}
\left(- \partial_i \partial_j - \delta_{ij} \partial_k \partial^k \right)
\left\{\left( \partial_k \partial^k\right)^{2n+\Delta} \right. \nonumber \\ 
& \left. \times \left(- \partial^i \partial^j A - \delta^{ij} \partial_k \partial^k A
+ \frac{1}{3}\partial^i \partial^j \partial_k \partial^k E
+ \frac{1}{3} \delta^{ij} \left( \partial_k \partial^k \right)^2 E \right)
\right\} \, .
\end{align}
By the variation of $E$, we obtain
\begin{align}
\label{Q56}
0 =& \partial_k \partial^k \left[ \frac{1}{8\kappa^2}
\left\{ \frac{4}{3}\left( - \partial_t^2 + \partial_k \partial^k\right)
\partial_k \partial^k E
+ \frac{4}{3} \partial_k \partial^k A
+ \frac{16}{9} \left( \partial_k \partial^k \right)^2 E \right\} \right. \nonumber \\ 
& + \frac{2^{2n-1+\Delta}}{3} \alpha U_0^{2n+\Delta}
\left(- \partial_i \partial_j - \delta_{ij} \partial_k \partial^k \right)
\left\{\left( \partial_k \partial^k\right)^{2n+\Delta} \right. \nonumber \\ 
& \left. \left. \times \left(- \partial^i \partial^j A - \delta^{ij} \partial_k \partial^k A
+ \frac{1}{3}\partial^i \partial^j \partial_k \partial^k E
+ \frac{1}{3} \delta^{ij} \left( \partial_k \partial^k \right)^2 E \right)
\right\} \right] \, .
\end{align}
Then we find $\partial_t^2 A = 0$ and therefore by using  
$A = \frac{1}{3} \partial_k \partial^k E$, we find $A=E=0$. 
Therefore we find all the scalar modes $\phi$,
$\lambda$, $h_{tt}$, $s$, $A$, and $E$ 
and all the vector modes $v_i$ and
$B_i$ do not propagate. 
The only propagating mode is massless graviton $C_{ij}$. 
This situation should be distinguished from that in the Ho\v{r}ava quantum gravity. 
The action for the massless graviton $C_{ij}$ is given by
\begin{align}
\label{Q58}
S =& \int d^4 x \left[ \frac{1}{8\kappa^2} \left\{
C_{ij} \left( - \partial_t^2 + \partial_k \partial^k\right) C^{ij}
\right\} \right. \nonumber \\ 
& \left. - 2^{2n-2+\Delta} \alpha U_0^{2n+\Delta} \left\{\left( \partial_k \partial^k
\right)^{n+1+\Delta}C_{ij}\right\}
\left\{ \left( \partial_k \partial^k \right)^{n+1} C^{ij} \right\} \right]
\, , 
\end{align}
which give the propagator of $C_{ij}$ as follows
\begin{align}
\label{Q60}
& \left< h_{ij}(p) h_{kl}(-p) \right>
= \left< C_{ij}(p) C_{kl}(-p) \right> \nonumber \\ 
& = \frac{1}{2} \left\{ \left( \delta_{ij} - \frac{p_i
p_j}{\bm{p}^2}\right)
\left( \delta_{kl} - \frac{p_k p_l}{\bm{p}^2}
\right) - \left( \delta_{ik} - \frac{p_i p_k}{\bm{p}^2}\right)
\left( \delta_{jl} - \frac{p_j p_l}{\bm{p}^2}
\right) \right. \nonumber \\ 
& \left. - \left( \delta_{il} - \frac{p_i p_l}{\bm{p}^2}\right)
\left( \delta_{jk} - \frac{p_j p_k}{\bm{p}^2}\right) \right\} 
\left( p^2 - 2^{2n+1+\Delta} \alpha \kappa^2 
U_0^{2n+\Delta} \bm{p}^{2 ( 2n+2+\Delta )} \right)^{-1} \, .
\end{align}
Here ${\bm{p}}^2 = \sum_{i=1}^3 \left( p^i \right)^2$ and  
$p^2 = - \left(p^0\right)^2 + {\bm{p}}^2$.
We should assume $\alpha < 0$ in order to avoid tachyon pole. 
In the ultraviolet region, the propagator behaves 
as $\sim$  $1/\left| \bm{p} \right|^4$ for $z=2$ ($n=0$) case
and $\sim$  $1/\left| \bm{p} \right|^6$ for $z=3$ ($n=0$) case 
and therefore the model could be power-counting renormalizable.
In the case of $z=2n +2$ ($n\geq 1$) or $z=2n+3$ ($n\geq 1$) case, 
the model could be (power-counting) super-renormalizable.

\section{FRW cosmology}

We now consider the FRW cosmology. 
We may start with a little bit general action: 
\begin{align}
\label{Q61}
&S = \int d^4 x \sqrt{-g} \left[
\frac{R}{2\kappa^2} - \sum_{\Delta=0,1} \sum_{n=0}^{n_\mathrm{max}} \alpha_n^\Delta \left\{
\left(\partial^\mu \phi \partial^\nu \phi \nabla_\mu \nabla_\nu - \partial_\mu \phi
\partial^\mu \phi \nabla^\rho \nabla_\rho \right)^n
P_\alpha^{\ \mu} P_\beta^{\ \nu} \right. \right. \nonumber \\ 
& \left. \times \left( R_{\mu\nu} - \frac{1}{2 U_0 }
\partial_\rho \phi \nabla^\rho \nabla_\mu \nabla_\nu \phi
\right) \right\} \Bigl\{ \left(\partial^\mu \phi \partial^\nu \phi
\nabla_\mu \nabla_\nu - \partial_\mu \phi \partial^\mu \phi
\nabla^\rho \nabla_\rho \right)^{n+\Delta}
P^{\alpha\mu} P^{\beta\nu} \nonumber \\ 
& \times \left. \left( R_{\mu\nu} - \frac{1}{2 U_0 }
\partial_\rho \phi \nabla^\rho \nabla_\mu \nabla_\nu \phi
\right) \right\} 
\left. - \lambda \left( \frac{1}{2} \partial_\mu \phi \partial^\mu \phi
+ U_0 \right) \right]\, .
\end{align}
In low energy, in addition to the Einstein-Hilbert term, 
$\Delta=0$, $n=0$ term in (\ref{Q61}) could dominate and the action (\ref{Q61}) could 
reduce to
\begin{align}
\label{Q62}
S \sim& \int d^4 x \sqrt{-g} \left[ \frac{R}{2\kappa^2} - \alpha_0^0
P_\alpha^{\ \mu} P_\beta^{\ \nu} \left( R_{\mu\nu} - \frac{1}{2 U_0 }
\partial_\rho \phi \nabla^\rho \nabla_\mu \nabla_\nu \phi \right) \right. \nonumber \\ 
& \times P^{\alpha\mu} P^{\beta\nu} \left( R_{\mu\nu} - \frac{1}{2 U_0 }
\partial_\rho \phi \nabla^\rho \nabla_\mu \nabla_\nu \phi
\right) 
\left. - \lambda \left( \frac{1}{2} \partial_\mu \phi \partial^\mu \phi
+ U_0 \right) \right]\, .
\end{align}
We now assume the FRW metric as 
$ds^2 = - {\rm e}^{2b(t)} dt^2 + a(t)^2 \sum_{i=1,2,3} \left(dx^i\right)^2$.
Then by the variation of $b$, we obtain the following FRW equation: 
$\frac{3}{\kappa^2} H^2 + 81 \alpha_0^0 H^4
= \rho_\mathrm{matter}$. 
Here we put $b=0$ after the variation and $\rho_\mathrm{matter}$ 
is the energy-density of the matter. 
We should note that the FRW equation is the first order differential equation 
with respect to the scale factor $a(t)$. 
If $\alpha_0^0<0$ and $\rho_\mathrm{matter}=0$, the FRW equation 
admits the de Sitter solution $H^2 = - \frac{1}{27 \alpha_0^0 \kappa^2}$, 
which may correspond to the inflation in the early universe. 


\section{Superluminal neutrinos}

The OPERA experiment results indicate towards the possibility that the
neutrino speed might exceed the speed of light\cite{:2011zb}. 
Motivated with the results, in Ref.~\refcite{Nojiri:2011ju}, a model of superluminal 
spinor by the spontaneous breakdown due to the Lagrange multiplier field has been 
proposed. 
The action we consider is
\begin{align}
\label{lvf1}
S = \int d^4 x \left[
\bar\psi \left\{ \gamma^\mu \partial_\mu
+ \alpha \left( P_\mu^{\ \nu} \gamma^\mu \partial_\nu\right)^{2n+1} \right\}
\psi - \lambda \left( \frac{1}{2}
\partial_\mu \phi \partial^\mu \phi
+ U_0 \right)\right] \, .
\end{align}
Here $\alpha$ is a constant, $n$ is an integer equal to or greater than $1$,
and $P_\mu^{\ \nu}$ is a projection operator again. 
By using (\ref{Q37}), the equation corresponding to the Dirac equation
$0=\left\{ \gamma^\mu \partial_\mu
+ \alpha \left( P_\mu^{\ \nu} \gamma^\mu \partial_\nu\right)^{2n+1} \right\} \psi$ 
looks as $0=\left\{ \gamma^0 \partial_0 + \gamma^i \partial_i
+ \alpha \left( \gamma^i \partial_i \right)^{2n+1} \right\} \psi$. 
Therefore the dispersion relation for the spinor is given by 
$\omega = k \sqrt{ 1 + \alpha^2 k^{4n} }$. 
Here $\omega$ is the angular frequency corresponding to
the energy and $k$ is the wave number corresponding to the momentum.
In the high energy region, the dispersion relation becomes
$\omega \sim |\alpha| k^{2n+1}$ 
and therefore the phase velocity $v_p$ and the group velocity $v_g$ are given, 
respectively, by 
$v_p \equiv \frac{\omega}{k} = |\alpha| k^{2n}$ and 
$v_g \equiv \frac{d\omega}{dk} = \left(2n+1\right) |\alpha| k^{2n}$, 
respectively.
When $k$ becomes larger, both  $v_p$ and $v_g$ become also larger
in an unbounded way and exceed the light speed.

\section{Summary}

In this report, we investigated the model with the spontaneous breakdown of 
the Lorentz symmetry by using the Lagrange multiplier field. 
We considered $F(R)$ gravity, power-counting renormalizable gravity, and 
superluminal spinor. 
We formulated the power-counting renormalizable gravity, and 
the superluminal spinor by using scalar projectors. 
For the power-counting renormalizable gravity, we have shown that 
the theory admits flat space solution and the only propagating mode is (higher
derivative) graviton, while scalar and vector modes do not propagate by 
developing the gauge-fixing formulation. 
We also gave a preliminary study of FRW cosmology indicates to
the possibility of inflationary universe solution.
The first FRW equation in the theory turns out to be the
first order differential equation which is quite unusual for higher derivative
gravity which normally leads to third order differential equation with respect
to scale factor.

\section*{Acknowledgments}

This report is based on the works written by the collaborations with S.D. Odintsov. 
The author also acknowledges J. Kluson for the collaboration in Ref.~\refcite{Kluson:2011rs}. 
This research has been supported in part
by MEC (Spain) project FIS2006-02842 and AGAUR(Catalonia) 2009SGR-994 (SDO),
by Global COE Program of Nagoya University (G07)
provided by the Ministry of Education, Culture, Sports, Science \&
Technology and by the JSPS Grant-in-Aid for Scientific Research (S) \# 22224003
and (C) \# 23540296 (SN).


\end{document}